# Impacts of shared autonomous vehicles: Tradeoff between parking demand reduction and congestion increase

Yusuke Kumakoshi[a], Hisatomo Hanabusa[b], Takashi Oguchi[c,*]

[a]*Research Center for Advanced Science and Technology, the University of Tokyo, Tokyo, Japan*
[b]*i-Transport Lab. Co. Ltd., Tokyo, Japan*
[c]*Institute of Industrial Science, The University of Tokyo, Tokyo, Japan*

**Abstract**

Shared autonomous vehicles (SAVs) can have significant impacts on the transport system and land use by replacing private vehicles. Sharing vehicles without drivers is expected to reduce parking demand, and as a side effect, increase congestion owing to the empty fleets made by SAVs picking up travelers and relocating. Although the impact may not be uniform over a region of interest owing to the heterogeneity of travel demand distribution and network configuration, few studies have debated such impact at a local scale, such as in transportation analysis zones (TAZs). To understand the impact in relation to geographical situations, this study aims to estimate the impacts of SAVs at the local scale by simulating their operation on a developed simulator. Using the mainland of Okinawa, Japan as a case study, it was found that parking demand was reduced the most in residence-dominant zones in terms of quantity and in office-dominant zones in terms of proportion. As a side effect of replacing private vehicles with SAVs, empty fleets increase congestion, particularly at the periphery of the city. Overall, the results show the heterogeneous impacts of the SAVs at the TAZ level on both land use and traffic, thus suggesting the importance of developing appropriate strategies for urban and transport planning when considering the characteristics of the zones.

*Corresponding author
Email: ykuma@cd.t.u-tokyo.ac.jp (Yusuke Kumakoshi),
hanabusa@i-transportlab.jp (Hisatomo Hanabusa),
takog@iis.u-tokyo.ac.jp (Takashi Oguchi)

# 1. Introduction

Shared autonomous vehicles (SAVs) can have a significant impact on a transport system. Technological advancements have enabled such vehicles to become an option for transport modes, approximated as fleets of driverless taxis. Society will likely accept the introduction of SAVs because the sharing services in various transport modes (Ferrero et al., 2018; Fishman et al., 2013; Younes et al., 2020) are increasing in popularity, and travelers can save costs by sharing the fabrication and maintenance cost of autonomous vehicles, in comparison to maintaining privately owned autonomous vehicles (Böosch et al., 2018; Litman, 2019). Although 100% use of SAVs may not be realistic today, both the benefits and detriments are worth studying as an upper limit of the potential impact of the new transport mode, and the results provide insight to practitioners (Fraedrich et al., 2019).

As a benefit, the replacement of private vehicles with SAVs will save parking space because it is unnecessary to park a vehicle after a trip is finished (Kondor et al., 2018; Zhang et al., 2015). In addition, designing an optimal layout of the parking facility for SAVs can further reduce the demand for parking spaces (Nourinejad et al., 2018). By repurposing the land that is made available for more beneficial uses, urban areas may gain vitality.

By contrast, congestion is expected to increase when SAVs are introduced, compared to the current situation with private vehicles, although the degree of increase depends on hypotheses regarding the SAV operation (Fagnant and Kockelman, 2014; Narayanan et al., 2020). Because a central control center is assumed to control the SAVs in general, the SAVs without travelers on board move to either pick up a waiting traveler or relocate themselves to the places where the expected travel demand is high. Thus, despite the efficiency of the vehicle operation, the empty fleet generated is considered as one reason for an increase in congestion (Vosooghi et al., 2019).

Faced with this change, transport planners and urban planners need to develop a new planning scheme for the age of SAVs. However, previous studies have limitations in providing recommendations to planners for the following three reasons. First, most of the simulated settings were unrealistic: either network or travel demand was hypothetically constructed (Kondor et al., 2018; Zhang et al., 2015), and the studied areas were insufficiently broad to consider the comprehensiveness of origin-destination (OD) trips (Okeke, 2020; Zhang and Guhathakurta, 2017). Second, a dynamic traffic flow was not considered when the parking demand was estimated (Okeke, 2020; Zhang and Guhathakurta, 2017; Zhang and Wang, 2020) This is problematic because the consequences of replacing private vehicles with SAVs cannot be quantitatively estimated, thus ignoring the impact of vehicles in the network that will park and operate. Finally, most studies have reported the performance of SAVs at the entire network level, not at a divided level, such as a road segment or a geographically delimited area called a transport analysis zone (TAZ). The heterogeneity of land use and the road network structure requires a high-resolution understanding regarding the results of introducing SAVs.

To understand the impact in relation to the geographical circumstances, this study aims to estimate the impact of SAVs on parking demand in an urban area, as well as that on traffic flow, using mesoscopic simulations on a real-world network and real OD data. A scenario for SAVs was developed, in which SAVs replace all current travel demand with private vehicles, and a control center (dispatcher) operates the SAVs. The operational strategy was designed based on strategies described in previous studies; therefore, the optimality of the strategy is beyond the scope of this paper.

More specifically, this paper attempts to answer the following questions using the real network of mainland Okinawa, Japan, as a case study:

- Which type of land use will reduce the parking demand the most under the SAV scenario compared to the private vehicle scenario?

- In which part of the road network will the congestion increase the most under the SAV scenario compared to the private vehicle scenario?

Although the simulation results may not be directly applicable to other cities, several implications generalized for transport and urban planning are discussed to infer the mechanism behind the observed phenomena.

The contributions of this paper are as follows:

1. A simulator of dynamic traffic flow was developed to estimate parking demand and traffic situations, applicable to both private vehicles and SAVs. This enables a realistic estimation of the impact of SAVs on both land use and traffic. In addition, the simulator may be applied to large-scale networks with a reasonable computational load.

2. The reduction in parking demand of SAVs compared to that of private vehicles is quantitatively estimated at the TAZ scale on mainland Okinawa, Japan. The reduction in terms of the number of parking slots was the largest in residence-dominant areas, while that in terms of proportion to the quantity of the original parking demand was the largest in the office-dominant area.

3. An increase in congestion by SAVs compared to private vehicles is also quantitatively estimated at the TAZ scale. The tradeoff between parking demand reduction and congestion increase at the TAZ scale was also demonstrated, and it provides evidence of the need to consider land use and traffic aspects when planners consider the introduction of SAVs.

## 2. Related studies

*2.1. SAVs: Operation and performance*

The operations of SAVs have been studied under various settings (Burns et al., 2013; Fagnant and Kockelman, 2014; Spieser et al., 2014).

They are typically modeled within the framework proposed by Levin et al. (2017), which distinguishes between two phases: matching and relocation. Matching is a phase in which vehicles are assigned to the travel demands. Optimization-based approaches (Hyland and Mahmassani, 2018; Ruch et al., 2018) and heuristic approaches (Bischoff and Maciejewski, 2016; Höorl et al., 2019) have been proposed, and the latter has often been employed in simulation studies because of the relatively light computational load. Relocation is a phase in which vacant vehicles reposition themselves to a place where the expected travel demand is high. Similar to the matching techniques, both optimization-based approaches (Pavone et al., 2012) and heuristic approaches (Fagnant and Kockelman, 2014; Höorl et al., 2019) have been proposed.

The performances of SAVs as a transport system have also been widely studied. In terms of benefits, a smaller energy consumption (Bauer et al., 2018), and reduced emissions (Jones and Leibowicz, 2019; Martinez and Viegas, 2017) have been argued, particularly with the application of electric vehicles; optimization of the fleets and scheduling of the vehicle charging is advocated for a greener shift (Jones and Leibowicz, 2019). The automated driving system also allows an increase in the traffic capacity of road networks by reducing the distance between running vehicles (Tientrakool et al., 2011). From the users' perspective, the accessibility of non-drivers to various activities may be enhanced (Meyer et al., 2017) because the SAVs can be used by people such as children and the elderly who do not possess a driver's license (Fagnant and Kockelman, 2015).

Meanwhile, an increase in vehicle kilometers traveled (VKT) is expected in general because of the empty fleets of SAVs picking up the next traveler and relocating themselves (Bischoff and Maciejewski, 2016; International Transport Forum, 2015; Kondor et al., 2018). The increased traffic demand may exceed the capacity of a road with a bottleneck, and therefore causes congestion. Consequently, the average travel speed in road sections with a bottleneck is lowered (Zhao and Kockelman, 2018), and the total travel and delay times are increased (Boesch et al., 2016; Böosch et al., 2018). At a regional scale, the heterogeneity of these changes over different TAZs has been pointed out, such as a traffic reduction in inner cities and an increase in suburban TAZs (Litman, 2019; Narayanan et al., 2020); however, a quantitative estimation with a real-world setting has yet to be conducted.

## 2.2. Parking demands of private vehicles and SAVs

Conventionally, parking demand at a zone scale has been estimated using the information on land use of the zones (Institute of Transportation Engineers, 2010); the scale is larger than individual parking slots and smaller than a city. Based on field surveys, the parking generation rate for each land use type was first inferred, and the rate was then multiplied by the area of the zone to obtain the estimation. A similar method has been adopted in Japanese cities (Ministry of Land, Infrastructure, Transport and Tourism, 2007), particularly to determine the number of parking slots needed at a large commercial facility, which aims at mitigating the congestion on the road around the facility, caused by vehicles cruising to find a parking spot (Shoup (2006)).

The impact of reducing the parking demand by introducing SAVs has been mostly estimated by simulation, owing to the complexity of the SAV behavior. Zhang et al. (2015) built an agent-based simulation model on a hypothetical grid city with randomly generated travel demand and found that up to 90% of parking demand can be reduced compared to that in a scenario without SAVs. Zhang and Guhathakurta (2017) extended the work to a real transport network and suggested that parking demand can be reduced by 4.5% with a penetration ratio of 5%. This work was further developed by Zhang and Wang (2020), who argued for future trajectories of parking demand at multiple time points. The International Transport Forum (2015, 2017a,b, 2018) also estimated on-street and off-street parking demand using their simulation tool.

However, these studies had the following limitations. First, simulation settings are often unrealistic. The travel demand or the network was virtually constructed, and the region of interest was limited to a neighborhood scale (Dia and Javanshour, 2017; Okeke, 2020). Second, a dynamic traffic flow was ignored: The link-level travel speed is often fixed for a certain time (Zhang et al., 2015; Zhang and Guhathakurta, 2017), or the travel time of an individual is derived from a hypothetical distribution (Kondor et al., 2018; Zhang and Wang, 2020). As Levin et al. (2017) discussed, ignorance of dynamic traffic flow leads to over-optimistic results. Third, most of the studies focused on the macroscopic performance of SAVs, namely, the total amount of parking demand over the study area (International Transport Forum, 2015); hence, knowledge at a microscopic scale such as each TAZ has not been sufficiently provided.

## 2.3. Traffic flow simulation

A traffic flow simulation has two active categories for practical use: microscopic and mesoscopic models. Microscopic models, such as VISSIM (Fellendorf and Vortisch, 2010) and Aimsun (Casas et al., 2010) reproduce the vehicle movement in detail, based on the car-following model. However, applying the model to large-scale road networks has difficulties in model calibration with dozens of driving behavior parameters, laborious data acquisition, and computational resource requirements (Horiguchi et al., 2000).

Mesoscopic models implement relatively simple flow models; the ease of parameter fitting, model building, and computation can be applied to large-scale studies. MATSim (Axhausen et al., 2016) is popular owing to its open-source policy (AMoDeus; Hörl et al., 2019), but the traffic flow model in MATSim does not sufficiently reproduce the physical queue on the roads (Mizokami et al., 2019). The simulation of urban road networks with a dynamic route choice (SOUND) (Yoshii and Kuwahara, 1995; i-Transport Lab, 2020) reproduces the queue evolution based on the simplified kinematic wave theory (Newell, 1993a,b) and variational theory (Daganzo, 2005), and applies to large-scale road networks (e.g.,

Tamamoto et al., 2004; Oshima et al., 2013; Oguchi et al., 2018). It deals with time-varying OD matrices of multiple classes in which the route choice characteristics are described as logit-type stochastic behavior dynamically considering the travel times, toll fees, road closures, and other factors. The authors consider SOUND to be more appropriate than other models to describe the spatio-temporal evolution of traffic congestion.

## 3. Methodology

### 3.1. Simulation framework

This section presents the simulation framework adopted in this study to estimate the impact of SAVs replacing all private vehicles. The simulation framework, based on that proposed by Levin et al. (2017), consists of two modules: one for an SAV dispatcher and the other for the traffic flow. Figure 1 describes the combination of these modules and the overall simulation framework. The dispatcher controls the SAVs through two phases: matching (Hörl et al., 2019) and relocation (Fagnant and Kockelman, 2014), both of which were implemented as sub-modules. The traffic flow module is implemented based on the dynamic traffic flow simulator SOUND. Because SOUND, by default, simulates the traffic flow from an OD matrix determined before the beginning of the simulation, the implementation includes a modification of SOUND that allows the time-varying OD matrices of the SAVs to be determined by the dispatcher module.

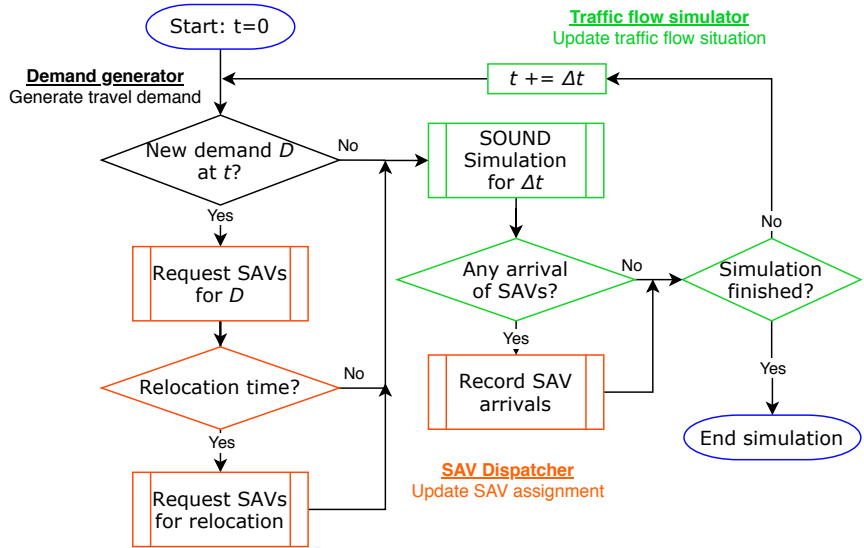

Figure 1 Simulation framework

### 3.2. Simulation scenarios

To compare the performance of the SAVs with that of private vehicles, two scenarios were developed: the current situation with private vehicles and the SAV scenario with 100% replacement of private vehicles with SAVs. Both scenarios used an identical network, transport analysis zones (called a zone hereafter), and the OD matrix of hourly travel demand. The former scenario served as the baseline for the comparison. In the latter, all SAVs were required to stop at SAV stations, which were placed one per zone. In addition, a car-sharing scheme was adopted, whereas a ride-sharing scheme was not, i.e., only one traveler could take a vehicle at a time. The simulation model is described in the following sections.

### 3.3. Model specification: SAV operation

The dispatcher in this study was assumed to have complete information on the SAVs and travel demand over the network (Levin et al., 2017). The initial distribution of the SAVs was determined proportionally to the expected total trip generation by zone (Eq. (1)):

$$SAV_{init,i} = |V|\frac{G_i}{\sum_{j \in Z} G_j} \quad (1)$$

where $SAV_{init,i}$ is the number of SAVs initially distributed to zone $i$, $|V|$ is the total number of SAVs, $G_i$ is the trip generation from zone $i$, and $Z$ is the set of zones.

During the simulation, the dispatcher module receives the vehicle time-varying location information from the traffic flow module. Depending on the individual vehicle locations, the module determines the matching and relocation strategies for each vehicle. The matching sub-module is called every 30 s, and the relocation sub-module is called every 5 min (Höorl et al., 2019). Travel demand is generated just before the matching sub-module is called, and the number of trips for each OD pair per 30 s is calculated from the hourly OD matrix. The hourly travel demand is uniformly attributed to each 30-s bin. If the number per bin is less than one, it is carried over the subsequent bin such that the number of trips is always an integer.

### 3.3.1. Matching

The matching sub-module first sorts these travelers in *first-come-first-serve* order in terms of their generated time. If no vehicle can arrive at an acceptable waiting time for a traveler, which is assumed to be identical to all travelers, the traveler is added to the waitlist of the next matching turn; the travelers in the waitlist are placed at the top of the sorted order. The dispatcher assigns sorted travelers to the available SAVs individually, identifying the closest vehicle for each traveler. The closeness is determined by the estimated time of arrival, calculated as the summation of the travel time of the links along the minimum travel-time route between the current position of the vehicle and traveler, and the time needed to drop off a traveler if the traveler is in the vehicle at the moment of assignment. The link travel time is updated at 10-min intervals; therefore, the estimated closeness of any pair of places changes every 10 min, reflecting the time-varying traffic situation.

### 3.3.2. Relocation

To efficiently apply the matching process, available vehicles are relocated to zones where the expected travel demand exceeds the supply of SAVs. The relocation strategy is determined using the *Block Balance* (Fagnant and Kockelman, 2014; Fagnant et al., 2015), which represents the balance between the expected travel demand for and SAV supply in the upcoming 5 min period in each zone. Positive and negative values indicate an oversupply and excessive demand, respectively. This value is defined as follows:

$$Block\ Balance = SAVs_{total} \left( \frac{SAVs_{Block}}{SAVs_{Total}} - \frac{Demand_{Block}}{Demand_{Total}} \right) \quad (2)$$

where $SAVs_{total}$ represents the total number of SAVs in the simulation, and $SAVs_{Block}$ represents the number of available SAVs in a given zone, namely, the number of parking (non-assigned) SAVs. In addition, $Demand_{Total}$ represents the expected total trip generation from all zones in the upcoming 5-min period, and $Demand_{Block}$ stands for the expected trip generation from the zone in the same period, both of which are calculated from the OD matrix of the study area. Zones with block balance values beyond the thresholds (-5 and +5, i.e., the same values as those in Fagnant et al. (2015)) send vehicles to or receive them from neighboring zones such that the value approaches zero.

### 3.4. Model specification: Parking demand estimation

### 3.4.1. Current situation

In this study, parking demand is defined as the number of individual parking slots needed, and the parking slot is distinguished from the parking space, which refers to the land occupied for parking use. The parking demand in the current situation with private vehicles was further classified into two categories based on the characteristic of the space: parking facilities and garages. The former is a facility that accommodates vehicles in the daytime (e.g., parking slots in a shopping mall), whereas the latter is the parking space required for keeping privately owned vehicles in current Japanese law. Note that both types of parking spaces are generally located outside the streets in Japan. Accordingly, this study does not consider the microscopic effects of vehicles in on-street parking on traffic flow, such as the reduction of the traffic capacity of the road.

The first estimation was conducted based on statistical data to estimate the total parking demand for each zone. The demand for parking facilities was calculated by multiplying the parking generation rate for Japanese cities (Ministry of Land and Tourism, 2018; Table 1) and the corresponding land use area in each zone, and that for the garages was estimated based on the proportional allocation of the registered number of private vehicles by the number of households per zone. These two values are added to each zone.

| Type | Rate [veh/m$^2$] | Source or justification |
|---|---|---|
| Office | 0.005 | Ministry of Land and Tourism (2018) |
| Commerce | 0.0067 | Ministry of Land and Tourism (2018) |
| Residence | 0 | Garage is dominant, and outside parking facilities are considered to be rare |
| Industry | 0.0011 | Ministry of Land and Tourism (2018) and employee data |
| Park | 0 | Parking slots for identifiable large facilities were manually added to the Ministry of Land and Tourism (2018) |
| Transport | 0.0005 | and employee data |
| Nature | 0 | The number of parking facilities in a natural area was regarded as negligible |

Table 1: Parking generation rate and the source by land use type used in the case study.

However, the first estimation may not consider the dynamic aspects of parking behavior, such as a turnover. To ensure that the estimated parking demand reflects this aspect, a second estimation was conducted by regressing the number of arriving trips in the morning peak hour by the first estimation of the parking demand over all zones in the study area. The estimated slope is the average turnover rate of parking spaces in the study area. After confirming the fit of these two variables, the final estimation of parking demand per zone was obtained by multiplying the number of arriving trips by the average turnover rate.

Under the SAV scenario, the parking demand was directly recorded at the SAV stations during the simulation. The maximum number of vehicles simultaneously parking in a station during the simulation period was regarded as the parking demand. The result of this approach is comparable to that of the current situation in that both approaches consider the travel behavior of the vehicles during the simulation.

To discuss the impact on a geographical scale, the space needed for one parking slot was assumed to be 30 m$^2$, i.e., 2.5 m in width and 6 m in length for a slot and 2.5 m × 6 m for the clearance space (Ministry of Land, Infrastructure, Transport and Tourism, 1992).

## 4. Case study of Okinawa mainland and the results

*4.1. Data preparation and experimental environment*

The authors conducted simulations for two scenarios: the current situation and the SAV scenario, using mainland Okinawa, Japan, as a case study. As the reason for this choice, all trips begin and end inside the region of interest; otherwise, trips whose origin or destination is outside the region may decrease the accuracy of traffic flow simulation, because such trips are not in the OD matrix. Under both scenarios, each traveler in the simulation was assumed to perform only one trip during the morning peak hour, and no trip chain was assumed. The study area was limited to the zones in Naha City and the surrounding zones (Figure 2), which form the core and periphery of the city.

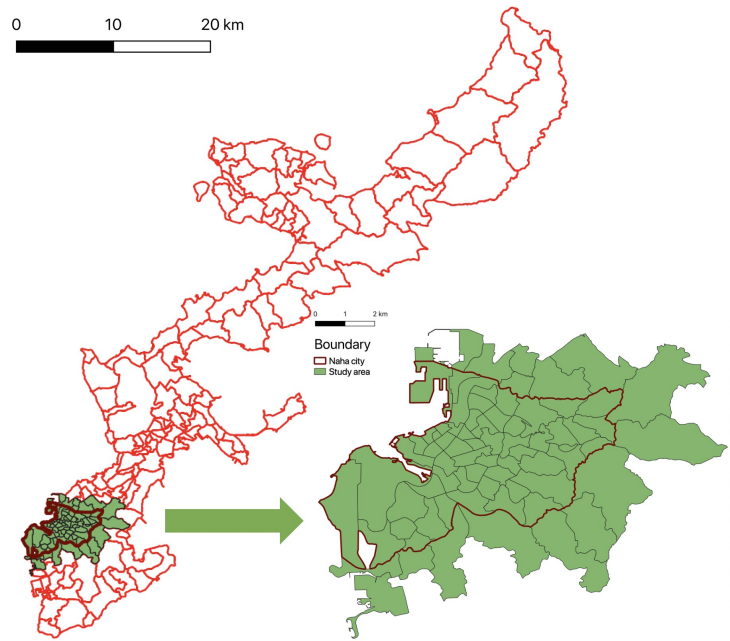

Figure 2: Location of the study area (Naha city, bottom-right) and the shape of the region of interest (Okinawa mainland, top-left).

The network data were provided by Digital Road Map (2015), the land use data were from Okinawa Prefecture, and the hourly travel demand OD matrices were constructed using the Road Census data (Ministry of Land, Infrastructure, Transport, and Tourism (2005b, 2015)) and an estimation method proposed by Kobayashi et al. (2012) (see Appendix A for a detailed explanation). The network consists of intra- and inter-urban roadways and has 34,663 nodes, 58,681 links, and 216 zones. The traffic light cycles and toll information for highways were also implemented during the simulation. A total of 225,505 trips were constructed over the morning peak hours (between 7:00 and 9:00 AM) for the entire network.

The computations were conducted on a 3.50-GHz Intel(R) Core(TM) i7-3970X CPU (implemented in C++), requiring approximately 20 min per scenario, and the parameters for the simulation, such as those concerning the route choice, road capacity, and others, were first calibrated to reproduce the actual situation in the real world. The simulation started at 4:00 AM, and the results between 7:00 and 9:00 AM were recorded for the analyses. Similarly, the entire network of the mainland was simulated, but only the results for the study area were used for the analyses. In addition, the fleet size was set to 60,000 after several preliminary simulation runs such that the average wait time of the travelers was at the same level as that in a previous study (Fagnant et al., 2015), which was approximately 1 min. This relatively short wait time ensures that the simulation setting is realistic from the travelers' perspective.

*4.2. Validation of the simulation*

The first parking demand estimation from land use data and the parking generation rates was regressed by the number of arrivals recorded in the OD matrix to derive the parking demand in this simulation model (Figure 3(a)). The estimated average turnover rate is approximately 1.70. The high goodness-of-fit ($R^2 = 0.79$) supports the validity of the adopted estimation method for parking demand. In effect, the estimated parking demand for Naha City was close to the statistical data, with an estimated 37,735 slots in the city, whereas the number of registered parking slots was 27,752 in 2005 (Ministry of Land, Infrastructure, Transport and Tourism (2005a)). Because these data excluded small parking facilities and garages, the actual number should be higher than the latter number, approaching the estimated value.

With regard to the traffic aspect, four elements were confirmed in the current situation with private vehicles. First, the constructed OD matrices reproduced the well-known peaks of the trip generation in the morning and evening (Figure 3(b)). Second, the estimated traffic flow at the link level was well aligned with the measured traffic flow in the Road Traffic Census, with $R^2 = 0.70$, as illustrated in Figure 3(c). Third, the length of the queuing vehicles in the morning peak hours reproduced a dynamic queue length change during congestion at the Naha Ohashi intersection, as shown in Figure 3(d). Finally, the estimated average travel speed of vehicles in Naha city (14.4 km/h) was close to the measured average travel speed (13.7 km/h) in the Road Traffic Census (2015). All of these elements validate the reproduction of the actual situation based on the simulation.

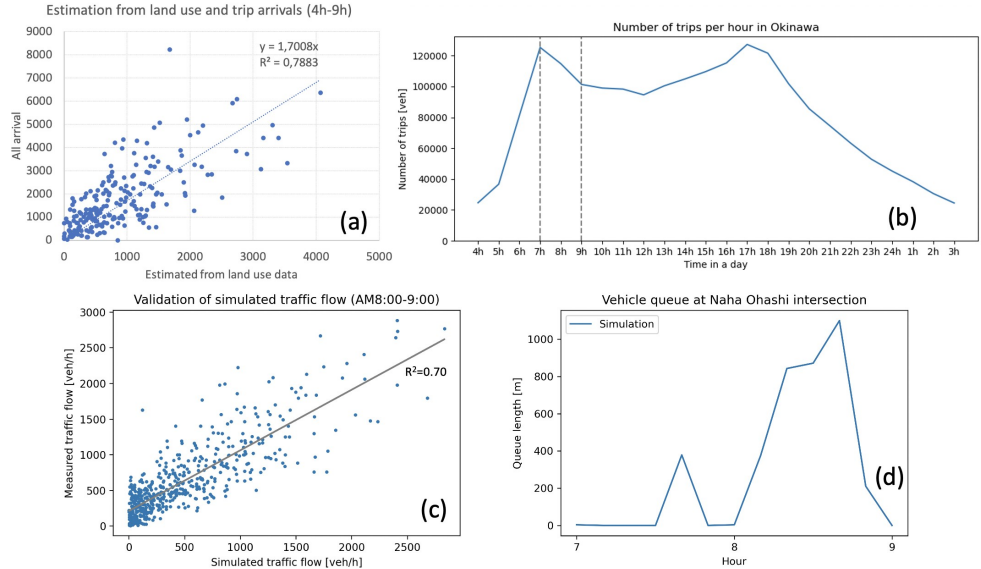

Figure 3: Data validating the simulation: (a) Number of arrival trips regressed by the static estimation of parking demand from land use and household data. Each point represents a zone, and the slope represents the average turnover ratio at the parking slots. (b) The number of trips over time in mainland Okinawa. (c) Correspondence of the estimated traffic flow to the measured value. Each point corresponds to a geographical point on the network. (d) Length of queueing vehicles at Naha Ohashi intersection.

## 4.3. Simulation results: comparison of Current situation and SAV scenario

This section presents the simulation results. Although each zone contained multiple land use types, for simplicity, this study characterizes a zone based on its most dominant land use type; for example, a residence-dominant zone refers to a zone where residential use is the most dominant.

### 4.3.1. Reduction in parking demand

The total parking demand in the study area significantly decreased in the SAV scenario compared with the current situation (94.0% reduction, from 409,625 in the current situation to 24,691 in the SAV scenario). Figure 4 shows the ratio of the reduction to the original parking demand (delta). Zones at the periphery of the city tend to reduce demand more than the average (94.0%), whereas those in the core of the city showed a rather moderate reduction (−90%–70%).

At the zone level, a parking demand reduction in terms of proportion to the demand in the current situation was the greatest in office-dominant zones (proportion of land occupied by parking space, from 44% to 5%), followed by residence-dominant zones (from 16% to 1%), as shown in Table 2. Although the demand was also reduced in the other three types of zones, the change was not drastic because the initial parking space occupied only a limited area from the beginning (5%–10%).

Regarding the absolute number of parking slots, the parking demand was reduced the most in the residence-dominant zones. Figure 5 (b) illustrates the proportion of the area (km$^2$) equivalent to the reduced demand to the total area of each dominant land-use type. In the office-dominant zones, although the proportion of land needed for parking largely decreased (−39%), the amount of space freed up was relatively small because the area of office-dominant zones was limited.

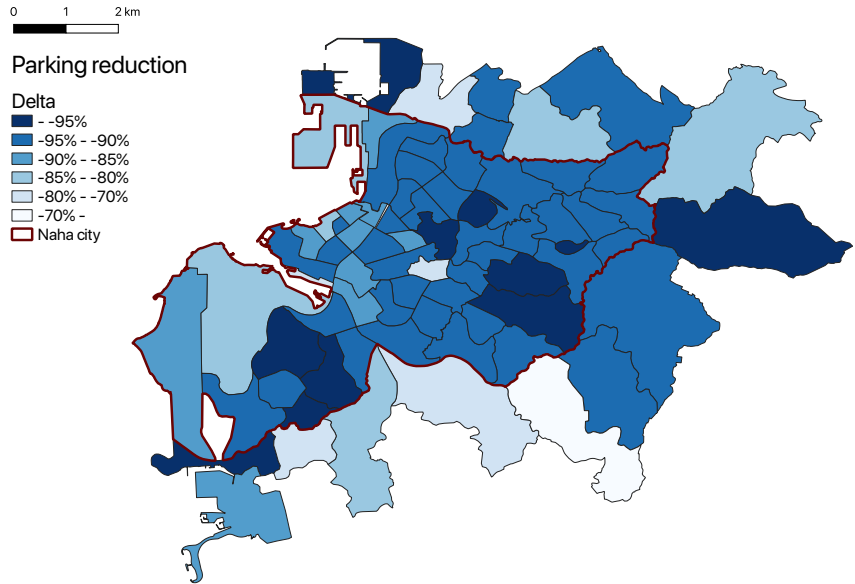

Figure 4: Reduction of parking demand in proportion (SAV–Current/Current)

| Zone type | Area [$km^2$] | Parking slot (ratio of parking space) | |
|---|---|---|---|
| | | Current | SAV |
| Office | 0.62 | 9 116 (44.4%) | 978 (4.8%) |
| Commerce | 8.32 | 32,090 (11.6%) | 2,877 (1.0%) |
| Residence | 63.2 | 346,337 (16.4%) | 18,541 (0.9%) |
| Industry | 0 | 0 (-) | 0 (-) |
| Park | 2.35 | 4,138 (5.3%) | 408 (0.5%) |
| Transport | 10.1 | 17,945 (5.3%) | 1,887 (0.6%) |
| Total | 84.7 | 409,625 (14.5%) | 24,691 (0.9%) |

Table 2: Parking demand reduction by land use type in the study area

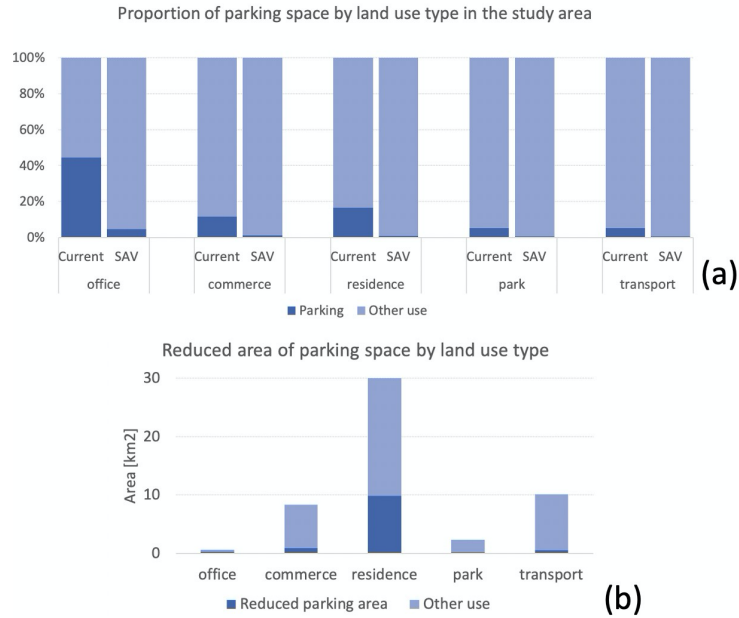

Figure 5: Proportion of parking space based on the scenario and land use type in Naha city

*4.3.2. Congestion increase*

Regarding the traffic flow, the larger amount of traffic demand of the SAVs than that in the current situation increased the VKT over the study area (+14.9%; Table 3). Consequently, the demand exceeded the traffic capacity at bottlenecks of the roads in the study area, thereby increasing the congestion. Under the SAV scenario, the total delay time increased by 32.5% and the average travel speed decreased by 9.2% compared to the current situation. The delay time in a road segment is defined as the difference between the actual travel time of a vehicle in the segment and the free flow travel time, and its increase indicates that vehicles spend more time on the road segment, leading to a slower average travel speed.

| Measurement in the study area | SAV | Current |
|---|---|---|
| VKT [veh × km] | 482,215 (+14.9%) | 419,785 |
| Delay time [veh × h] | 22,010 (+32.5%) | 16,606 |
| Average travel speed [km/h] | 14.0 (−9.2%) | 15.4 |

Table 3: Comparison of traffic flow status per scenario by vehicle kilometers traveled (VKT), delay time, and average travel speed.

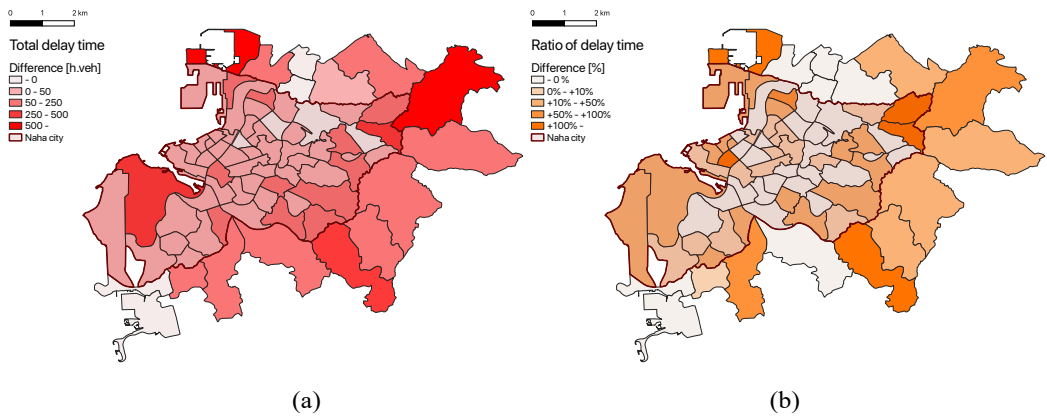

(a) (b)

Figure 6: (a) The total delay time in vehicle hours per zone and the (b) proportion of the delay time over the free flow time.

At the zone level, an increase in the delay time was observed, particularly in the zones at the periphery of the city. The increase in the total delay time per zone in Figure 6 (a) may have been biased by the difference in the total length of the road segments in each zone: The longer a segment is, the longer the time the vehicle needs to run through the segment, and the delay time itself may increase simply because of the longer segment length. Considering this possibility, the increase in the ratio of the total delay time to the total free flow time per zone is illustrated in Figure 6(b), and the high ratio at the periphery of the city provides solid evidence for the increase in congestion.

### 4.3.3. Tradeoff between parking demand reduction and congestion increase

To understand the tradeoff between parking demand reduction and congestion increase, the correlations between them were examined. To represent the congestion, the VKT, delay time, and average travel speed were chosen as the variables, and the difference in each variable between the two scenarios was used, i.e., the values in the current situation were subtracted from those in the SAV scenario. The correlation between these variables and the difference in parking demand, calculated in the same way as explained, was then calculated. The VKT was included because it provides complementary information to the other two, although an increase in the VKT itself does not indicate an increase in congestion.

Figure 7 illustrates the Spearman's rank correlation matrix with all values $p < 0.05$, indicating that the difference in parking demand is weakly correlated with the difference in the VKT (Spearman's rho = −0.145), moderately correlated with the delay time difference (rho = −0.426), and barely correlated with average difference in travel speed (rho = 0.045). The moderate correlation between the parking demand reduction and the delay time increase suggests that the replacement of private vehicles with SAVs, and consequently the pick-up and relation movements have an impact on the parking demand and the delay time at the zone level. By contrast, the small degree of correlations concerning the VKT, and the average travel speed implies that other elements such as the road network configuration influence the variance of the two variables.

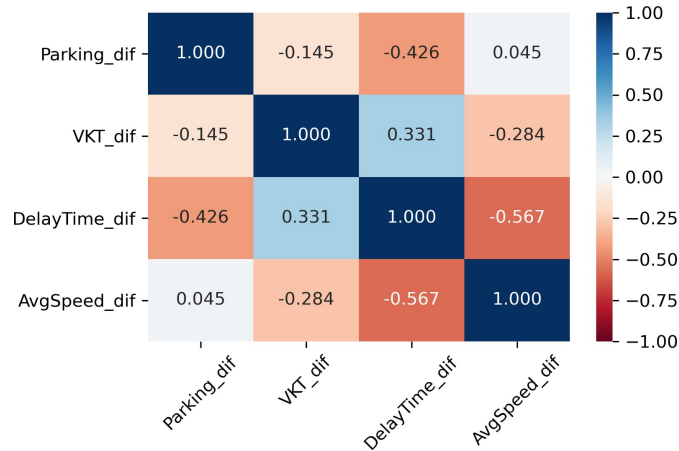

Figure 7: Correlation matrix of the differences of parking demand, VKT, delay time, and average travel speed under two scenarios

## 5. Discussion

### 5.1. Reasons for the parking demand reduction

This section discusses the possible reasons for parking demand reduction. One reason is the smaller fleet size of the SAVs compared to that of the private vehicles in the current situation. According to the Ministry of Land, Infrastructure, Transport and Tourism (2015), the total number of vehicles owned by the households in mainland Okinawa was 323,131, whereas that of the SAVs simulated in this study was 60,000. Assuming that one vehicle requires one parking slot, the estimated reduction (263,131 vehicles) accounted for 68.4% of the total parking demand reduction in terms of the number of parking slots (Figure 8). This is consistent with the result indicating that the reduction in parking demand in terms of the number of parking slots was the largest in the residence-dominant zones, where most of the garages are located.

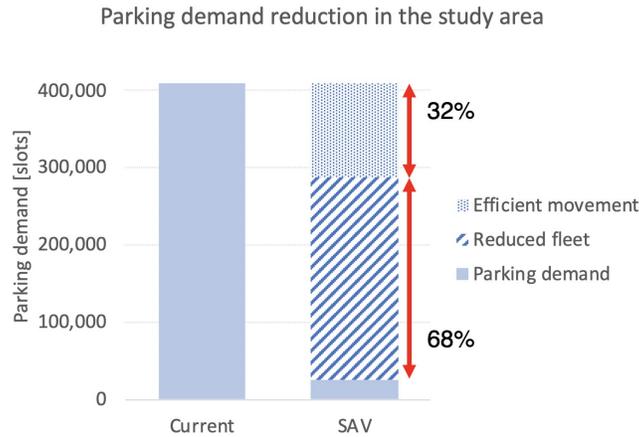

Figure 8: Decomposition of the reduced parking demand in mainland Okinawa

The remaining part of the reduction (31.6%) can be explained by the efficient movement of the SAVs, i.e., because the SAVs serve travelers consecutively, they spend less time in parking spaces compared to private vehicles, thus reducing the parking demand. Such a movement is enabled by the temporally distributed travel demands such that the dispatcher can find travelers wishing to depart after an SAV finishes its trip. In addition, the spatial distribution of travel demands affected the degree of reduction. If the OD pairs of the demand were more dispersed than those during the simulation, the size of the empty fleet of SAVs would increase and the SAVs would spend less time in the parking space than the computed SAV scenario, resulting in a larger reduction in parking demand. Therefore, the reduction in parking demand depends on the OD pattern in the study area.

*5.2. Reasons for increased congestion*

Consistent with previous studies (Fagnant and Kockelman, 2018; Levin et al., 2017; Martinez and Viegas, 2017), the total delay time and the VKT in the SAV scenario increased by 32.5% and 14.9%, respectively, compared to the current situation. Because the larger traffic volume of the SAVs as approximated by the VKT in comparison to that of the private vehicles increased the total delay time, it is natural to argue the reasons for the increase in the VKT.

The first reason is the empty fleet of SAVs (Vosooghi et al.,2019). Because both scenarios used an identical OD matrix of the travelers, the amount of the VKT recorded by SAVs with travelers is no less than that of the current situation. Hence, the empty fleet is an additional part of the total VKT under the SAV scenario. The amount of this part also depends on the OD pattern in the study area. If the OD pairs were more spatially dispersed and less temporally distributed than the simulated scenarios, the SAVs would run more on the roads, thus adding further VKT with the empty fleet.

A second reason is the detour behavior of the SAVs, which has not been discussed in previous studies. Suppose an SAV with a traveler trying to take the same route (sequence of road segments) as a traveler takes under the current situation. With the SAV scenario, the route will take more time than it originally did in the current situation because of the congestion primarily caused by the empty fleets of the other SAVs. Because of this congestion, the dispatcher may seek another route for the SAV that allows it to arrive at the destination faster than the original route, even if the total distance of the alternative route is longer than that of the original route. This type of detour behavior would have caused the increase in the VKT, probably increasing the congestion on roads where the capacity was low.

The combination of the two reasons led to the overall increase in the VKT and thus the congestion. However, the extent of the increase differs from zone to zone; for example, the total delay time increased more in the zones at the periphery of the city than in the core of the city. A reason for this heterogeneity is the relative difficulty of matching between the travel demands and SAVs. Because the SAV stations at the periphery were more sparsely located than those in the core, SAVs picking up travelers and waiting at those stations must run longer than in the case between two zones in the core of the city, thereby causing congestion. In effect, the traffic volume of the empty fleet was observed more at the periphery than at the core (see Appendix Figure B.9).

### 5.3. Impact of reduction in parking demand through repurposing

This section provides an interpretation of the parking demand reduction in terms of the activities of citizens, focusing on three example zones (Table 4). The parking demand in spatial terms under the SAV scenario was reduced from the current situation by 29.1% in a commerce-dominant zone (zone 14). If all space freed up in the zone is repurposed for commercial use, the floor space of the commercial area in this zone is estimated to increase by 35%, considering the presence of multi-story buildings in the zone (Appendix C details the calculation procedure). Such repurposing contributes to the economic vitalization of the zone. In the office-dominant zone (zone 214), the proportion of space freed up to the total area was 38.1%. Such space can be repurposed, for example, for stockpiling goods for contingencies, which may enhance the resilience of the area. The quality of life of the citizens may also be enhanced, particularly in a residence-dominant zone (zone 25) where 18% of the area of the zone can be freed up. Because most of the space freed up is garage space, typically located on privately owned land, the space provides opportunities for various activities (Gehl, 2013). All of these interpretations confirm that a parking demand reduction under the SAV scenario has a significant impact on the lives of citizens.

| Zone | Main usage | Total area [m$^2$] | Area of main usage [m$^2$] | Parking space Current [m$^2$] | Parking space SAV [m$^2$] | Space freed up [%] |
|---|---|---|---|---|---|---|
| 14 | Commerce | 330 708 | 135 819 (43.2%) | 106 350 (32.2%) | 10 230 (3.1%) | (29.1%) |
| 214 | Office | 167 989 | 31 865 (78.1%) | 72 600 (43.2%) | 8 550 (5.1%) | (38.1%) |
| 25 | Residence | 822 958 | 614 705 (77.8%) | 157 650 (19.2%) | 8 160 (1.0%) | (18.2%) |

*Percentages between parentheses are ratios of the value over the total area in a zone.

Table 4: Examples of parking demand reduction by zone

*5.4. Limitations*

This study had three main limitations. First, although the case study provided insight into the mechanism behind the parking demand reduction and increase in congestion, it should be further studied whether the implications apply to other cities with different network configurations and travel demand patterns. The optimization of the number and distribution of SAV stations, total fleet size, and distribution of fleets with riding fee settings should also be investigated. Second, assuming the highest parking demand during a 24-h period, this paper studied only the morning peak hours, because a 24-h simulation study is not easy to conduct without trip chain information. Third, a scenario with a mixture of transport modes, such as private vehicles and SAVs, will provide implications for a transitional phase from the current situation. Although this paper argued the upper limit of the potential impact of replacing private vehicles with100% SAVs, scenarios with other penetration rates need to be specified. Extending the simulation presented in this paper with multiple vehicle classes will contribute to this point.

## 6. Conclusion

By studying the impact of SAVs on parking demand and traffic flow, this study found an intrinsic connection between the land use and transport system: A drastic change by the new transport mode brings about a considerable reduction in the parking demand and an increase in traffic flow because the vehicles parked in the current situation tend to run under the SAV scenario. Moreover, if the parking space is repurposed, the land cell will attract more travel demand, thus causing a further increase in traffic flow. Therefore, urban planners and transport planners should collaborate, recognize mutual dependency, and achieve a better urban environment. In the course of planning, society must reach an agreement on the balance between the benefits from the parking demand reduction and the detriments from the congestion increase. This study demonstrated two upper limits in a 100% SAV scenario, namely, a 94.0% reduction in parking demand and a 32.5% increase in the total delay time compared to the current situation, and a scenario with a lower penetration rate of SAVs will result in smaller changes. Each society may have a point of agreement depending on the value of time, and the quantitative information from the simulations provides evidence on such social decision-making.

If a majority of the citizens regard the degree of the detriments unacceptable, the transport planners must reorganize the transport system to reduce congestion, such as reinforcing the traffic capacity at intersections, increasing the number of SAVs during service, and allowing ridesharing in the SAVs to reduce the empty fleet. Urban planners, in turn, are asked to reorganize the land, including the freed up space, when considering the induced travel demand after the reorganization. It is, therefore, necessary to further develop such a planning scheme in the coming years.

**Acknowledgment**

The authors are grateful to the Okinawa General Bureau, Cabinet Office, and Okinawa Prefecture Land Policy Division for data acquisition for the case study. We thank Prof. Manfred Boltze (Technische Universität Darmstadt), who inspired this paper, Prof. Hironori Kato (University of Tokyo) for the insightful and constructive comments, and Dr. Ryota Horiguchi (i-Transport Lab) for the technical support.

**Appendix A. OD estimation from traffic flow data**

The OD matrix estimation method (Kobayashi et al., 2012) took the initial daily OD data as input, which was based on the Road Traffic Census in 2005, and calibrated the parameters of the simulation such that the traffic flow at the road segment level was close to the actual data from 2015 (Ministry of Land, Infrastructure, Transport, and Tourism, 2015). To infer the OD traffic flow for every hour from the daily OD matrix, time coefficients were determined based on the temporal distribution of traffic flow on National Route 58 in Naha city, which has the largest daily traffic volume in Okinawa. By calculating the estimated traffic volume in the daily OD matrix proportionately to the time coefficients, hourly OD matrices were obtained.

**Appendix B. Movement of SAVs**

Figure B.9 illustrates the traffic flow generated for pick-up and relocation from 8:00 to 9:00 AM, during which the road is more congested than the preceding hours. At this time, the empty fleet is present more at the periphery of the city rather than in the core of the city.

**Appendix C. Estimation of the floor space under repurposing**

It was assumed that all of the space freed up in the SAV scenario owing to the parking demand reduction was repurposed for commercial use, which occupied 43.2% of the land in zone 14 (136,000 $m^2$). This space amounts to 29.1% of the land (96,120 $m^2$ = 106,350–10,230), and this area was converted into commercial floor space for interpretation. The average floor area ratio of the zone was set to 200%, which is half of the maximum floor area ratio of the zone (400%), which was necessary because not all buildings in the zone reached the maximum floor area ratio, according to Okinawa Prefecture. Therefore, the space freed up (96,120 m$^2$) is equivalent to 35.3% of the current commercial floor space (96,120 m$^2$/(136,000 m$^2$ × 200%)).

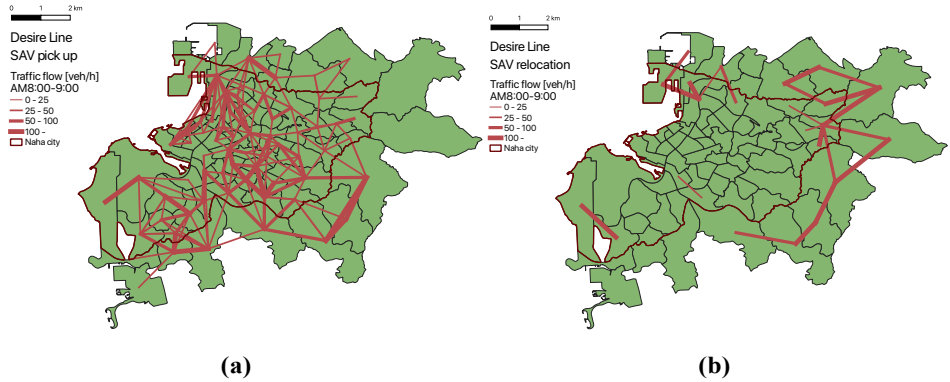

(a) (b)

Figure B.9: Desired line of SAV from 8:00 to 9:00 AM. (a) Pick up and (b) relocation. Both ends of the lines represent the OD pair, and the width of the lines corresponds to the volume of the traffic flow.